\documentclass[a4paper]{spie}

\usepackage{graphicx}
\usepackage{dcolumn}
\usepackage{bm}
\usepackage{amsmath,epsfig}

\title{Correlation filtering in financial time series}

\author{T. Aste\supit{a}, T. Di Matteo\supit{a}
\skiplinehalf
\supit{a}Department of Applied Mathematics, Research School of Physical Sciences, The Australian National University, 0200 Canberra, ACT, Australia.
\skiplinehalf
 M. Tumminello\supit{b}, R. N. Mantegna\supit{b}
\skiplinehalf
\supit{b}INFM Unit\`a di Palermo and Dipartimento di Fisica e Tecnologie Relative, 
Universit\`{a} di Palermo, Viale delle Scienze, Palermo, I-90128, Italy.\\
}

\authorinfo{
Corresponding author: Tiziana Di Matteo, 
e-mail: tiziana.dimatteo@anu.edu.au; 
phone 61 2 6125 0166; 
fax 61 2 6125 0732 \\
\rightline{
\begin{minipage} {16.cm}
\large Published in {\it Noise and Fluctuations in Econophysics and Finance}, Edited by D. Abbott, J.-P. Bouchaud, X. Gabaix, J. L. McCauley, Proc. of SPIE, Vol. 5848 (SPIE, Bellingham, WA, 2005) 100-109. (Invited Paper)
\end{minipage} }
}

\begin{document}

\maketitle

\date{\today}
\begin{abstract}
We apply a method to filter relevant information from the correlation coefficient matrix by extracting a network of relevant interactions. 
This method succeeds to generate networks with the same hierarchical structure of the Minimum Spanning Tree but containing a larger amount of links resulting in a richer network topology allowing loops and cliques. In Tumminello et al.~\cite{TumminielloPNAS05}, we have shown that this method, applied to a financial portfolio of $100$ stocks in the USA equity markets, is pretty efficient in filtering relevant information about the clustering of the system and its hierarchical structure both on the whole system and within each cluster. In particular, we have found that triangular loops and 4 element cliques have important and significant relations with the market structure and properties. Here we apply this filtering procedure to the analysis of correlation in two different kind of interest rate time series (16 Eurodollars and 34 US interest rates).
\end{abstract}

%

\section{Introduction}
The collective behavior of a system comprised of many elements is well described by the matrix of correlation coefficient among the elements.
The main difficulty in the analysis of such correlations arises from the fact that the information in the whole correlation matrix is often huge, containing correlation coefficients for all the pairs of elements in the system.
The challenge is to extract the smallest sub-set of such correlations which is able to describe accurately the collective behavior of the whole system. 
From a topological perspective, the correlation matrix can be represented by the complete graph (each node connected with all the other nodes) with edges weighted by the value of the correlation coefficient.
The problem of extracting a system of meaningful interactions is now translated into the problem of reducing the complete graph into a sub-graph which keeps only the edges which best describe the system of interactions.
Such a sub-graph must contain the maximum amount of information about the system's collective behavior while keeping the simplest possible structure.
It was shown in Ref.~\cite{Mantegna99} that a method to investigate correlations in financial systems consists in extracting a minimal set of relevant interactions associated with the strongest correlations belonging to the Minimum Spanning Tree (MST).
The MST maintains only the minimum number of links necessary to connect the network ($n-1$ links for a network with $n$ nodes) and, by construction, it keeps the links associated with the strongest correlations.  
This method turns out to be very effective in revealing the hierarchical structure contained in the correlation matrix.
However, the reduction to a minimal skeleton of links is necessarily very drastic in filtering correlation based networks loosing therefore valuable information. 
The necessity of a less drastic filtering procedure has already been raised in the literature. 
For example, an extension from trees to more general graphs generated by selecting the most correlated links, up to a given threshold, was proposed by Onnela et al.~\cite{Onnela03}. 
However, this method depends on the choice of the threshold and, when the number of kept links is constrained to be of the same order of the number of nodes (as for the MST), this method tends to generate disconnected networks made of several isolated, highly connected, sub graphs.
On the contrary, we seek for a connected network with no isolated nodes as in the case of the MST.

Since the MST method has been proved to be very effective in extracting hierarchies from the correlation matrix, its extension to a richer graph must be constructed in a way to maintain such a hierarchical skeleton while including a larger amount of links. 
A solution to this problem was recently proposed by introducing a new technique which enables to produce networks with tunable information content~\cite{Noi,TumminielloPNAS05}.
This method is based on the idea of connecting iteratively the most correlated nodes while constraining the resulting network to be embedded on a given surface.

We here apply this filtering procedure to the correlation matrix of \emph{16 Eurodollar Interest Rates}~\cite{DiMatteo} and \emph{$34$ different kinds of interest rates in money and capital markets}~\cite{DiMatteoMant,DiMatteoAless04}. 
We investigate the data clustering and differentiations seeking for basic sub-structures and exploring their hierarchical gathering and growth.

The paper is organized as following: in Section~\ref{Embedding}, we present the general idea regarding the embedding of the complete graph on hyperbolic surfaces. In Section~\ref{Corrnet}, we discuss the case of planar embeddings by considering the Planar Maximally Filtered Graph (PMFG) introduced in Ref.~\cite{Noi,TumminielloPNAS05}  and we compare it with the MST. Section~\ref{filtering} shows results for different sets of interest rates and in Section~\ref{clique} the emergent clique structures are analyzed and discussed. Finally the conclusions are given in Section~\ref{conclusion}.

\section{Embedding networks on hyperbolic surfaces}
\label{Embedding}

Any orientable surface can be topologically classified in terms of its {\it genus} which is the largest number of non-intersecting simple closed cuts that can be made on the surface without disconnecting a portion. The genus $g$ is a good measure of complexity for a surface: under such a classification, the sphere ($g=0$) is the simplest surface; the torus is the second-simpler ($g=1$); etc.
To a given network a genus can always be assigned. It is defined to be equal to the minimum number of handles that must be added to the plane to \emph{embed} the graph without edge-crossings.

Consider a system of $n$ interacting elements which are collectively fluctuating in a stochastic way.
In first place, these mutually-related stochastic variables can be considered all connected to each other in an $n$-th order complete graph ($K_n$).  
A \emph{weight} can associated to each link and a natural choice for the weight of a link between node $i$ and node $j$ is the correlation coefficient $c_{i,j}$.
The challenge is to locally simplify the network by keeping only the most significant interactions (largest correlations) and simultaneously extracting global information about the hierarchical organization of the whole system.
Hereafter we show that a very powerful way to proceed is to map the complete graph into a significant sub network on a 2-dimensional (2D) hyperbolic surface.
This approach has several attractive features: 
1) it provides new measures to characterize complexity; 
2) it gives a locally-planar representation;
3) it provides a hierarchical ensemble classification;
4) it allows the application of topologically invariant elementary moves~\cite{p16,p17,AsteSherr,Noi}.
In addition, let us stress that \emph{any} network can be embedded on a surface, therefore this approach is completely general and the genus of the embedding surface acts as a constraint on the complexity of the network.
Indeed, Ringel and Youngs have shown that an embedding of $K_n$ is always possible in an orientable surface $S_g$ of genus~\cite{Ringel1968}
\begin{equation}
\label{g*}
g \ge g^* = \lceil{ \frac{(n-3)(n-4)}{12} }\rceil
\end{equation}
 (for $n \ge 3$ and with $\lceil x \rceil$ the ceiling function  which returns the smallest integer number $\ge x$).
Therefore, providing a sufficiently high genus $g \ge g^*$, we can always generate embeddings that contain the whole information present in the complete graph.
Such a reduction to 2D cannot be in general implemented simply on the plane. 
Indeed, the complexity of the surface will increase with the complexity of the graph and multi-handled hyperbolic surfaces must be used (for $n > 7$ one has $g^* >1$, and hyperbolic surfaces are needed to embed the complete graph).

The embedding of the complete graph on $S_g$ is locally planar but this local simplification has been achieved at the expenses of an high complexity in the global surface ($g^*$ scales with $n^2$).
A reduction to simpler surfaces with lower genus is therefore necessary.
This can be done only by removing some of the ${n(n-1)}/{2}$ links of $K_n$ transforming the complete graph into a less connected network.
For instance, the opposite extreme to $K_n$ is the spanning tree which is the connected graph with minimum number of links ($n-1$) and it can be embedded on a surface of genus $g=0$ (the sphere).
In the next section we discuss an algorithm~\cite{TumminielloPNAS05} which allows to generate networks embedded on surfaces with arbitrary genus and we discuss in details the case $g=0$.

\section{From correlations to networks}
\label{Corrnet}

Starting from a correlation coefficient matrix, we construct a graph G embedded on a surface $S_g$ of genus $g$ by means of the following procedure: from a set of disconnected elements, by following an ordered list of pair of elements sorted in decreasing order of the correlation coefficient between two elements, we connect two elements if and only if the resulting network can still be embedded on $S_g$. Elsewhere the two elements are left disconnected and the subsequent pair in the list is considered.
It is important to stress out that this construction is analogous to the procedure which generates the MST.
Indeed, the only difference between the two procedures is that in the case of MST a link is inserted if and only if the network after such connection is a forest or a tree. 

It has been proved in Tumminello et al.~\cite{TumminielloPNAS05} that \emph{at any step of construction of the MST and graph G of genus $g$ if two elements are connected via at least one path in one of the considered graphs then they are connected also in the other one}.
This fact implies that the MST is always contained in G. Moreover, this also implies that the formation of connected clusters of nodes during the construction of a graph G coincides with the formation of the same clusters in the MST at the same stage of the construction.
In other words the hierarchical structure associated to G coincides with the one of the MST \emph{at any stage of the construction}.

The graph G of genus $g$ is the most connected graph for a given embedding $S_g$ and number of nodes $n$. The graph G is a triangulation of $S_g$. It has $3(n-2+2g)$ (for $g \leq g^*$) links and the addition of a further link is impossible without edge-crossings or the increase of the surface genus. A special case is the PMFG~\cite{TumminielloPNAS05} which corresponds to $g=0$, i.e. to a planar embedding~\cite{Planar}. This is the simplest embedding and coincides with the one for the MST. In this respect, the MST and the PMFG are the two extreme cases having the same topological complexity but having the minimum ($n-1$) and the maximum ($3n -6$) number of links respectively.

The main structural difference between the PMFG and the MST is that the PMFG allows the existence of loops and cliques. 
A clique of $r$ elements ($r$-clique) is a complete subgraph that links all $r$ elements.
Only cliques of 3 and 4 elements are allowed in the PMFG. Indeed, topological constraints expressed by the Kuratowski's theorem~\cite{Planar} do not allow cliques with a number of elements larger than 4 in a planar graph. 
Larger cliques can only be present in graphs with genus $g>0$ and the larger the value of $g$ the larger is the number of elements $r$ of the maximal allowed clique (specifically $r \le \frac{7 + \sqrt{1+48 g}}{2}$)~\cite{Ringel}.

\begin{figure}
\begin{center}
\includegraphics[width = 0.5 \textwidth]{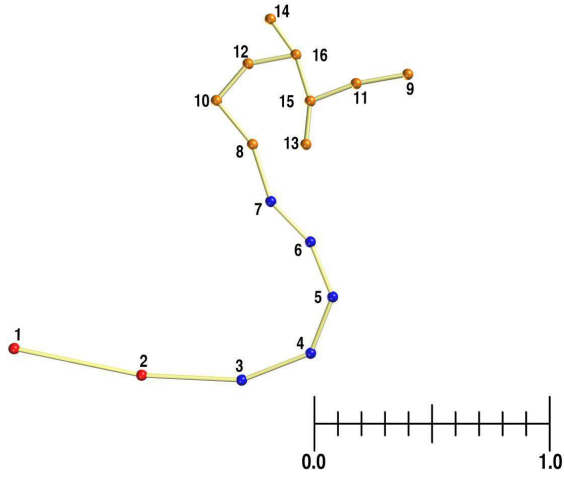}
\end{center}
\caption{Two dimensional representation of the minimum spanning tree for the $16$ Eurodollar interest rates. 
The edge-lengths are equal to the metric distances $d_{i,j}=\sqrt{2(1-c_{i,j})}$. The labels correspond to the maturity dates expressed in units of 3 months.}
\label{f.MSTEurod}
\end{figure}

\begin{figure}
\begin{center}
\includegraphics[width = 0.5 \textwidth]{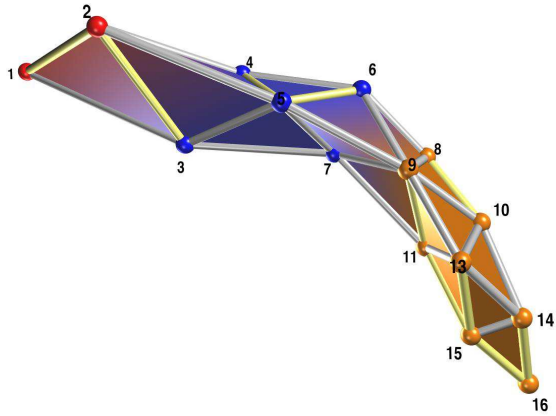}
\end{center}
\caption{ Three dimensional representation of the embedding on $S_0$ of the PMFG of the correlation structure for the $16$ Eurodollar interest rates. 
Each edge-length corresponds to the metric distance $d_{i,j}=\sqrt{2(1-c_{i,j})}$.
The labels correspond to the maturity dates expressed in units of 3 months.}
\label{f.3DEurodollar}
\end{figure}

\section{Correlation filtering in financial data}
\label{filtering} 

In the previous section we have introduced a \textit{general} method for constructing a network of genus $g$ from the correlation matrix. 
Here we constraint ourselves to the case $g=0$, i.e. to the PMFG and we present two examples of PMFG graphs obtained by means of such correlation based procedure. 
The first example concerns a set of 16 Eurodollars Interest rates~\cite{DiMatteo}.
The second example concerns a set of 34 US interest rates~\cite{DiMatteoMant}.

\begin{figure}
\begin{center}
\includegraphics[width = 0.5 \textwidth]{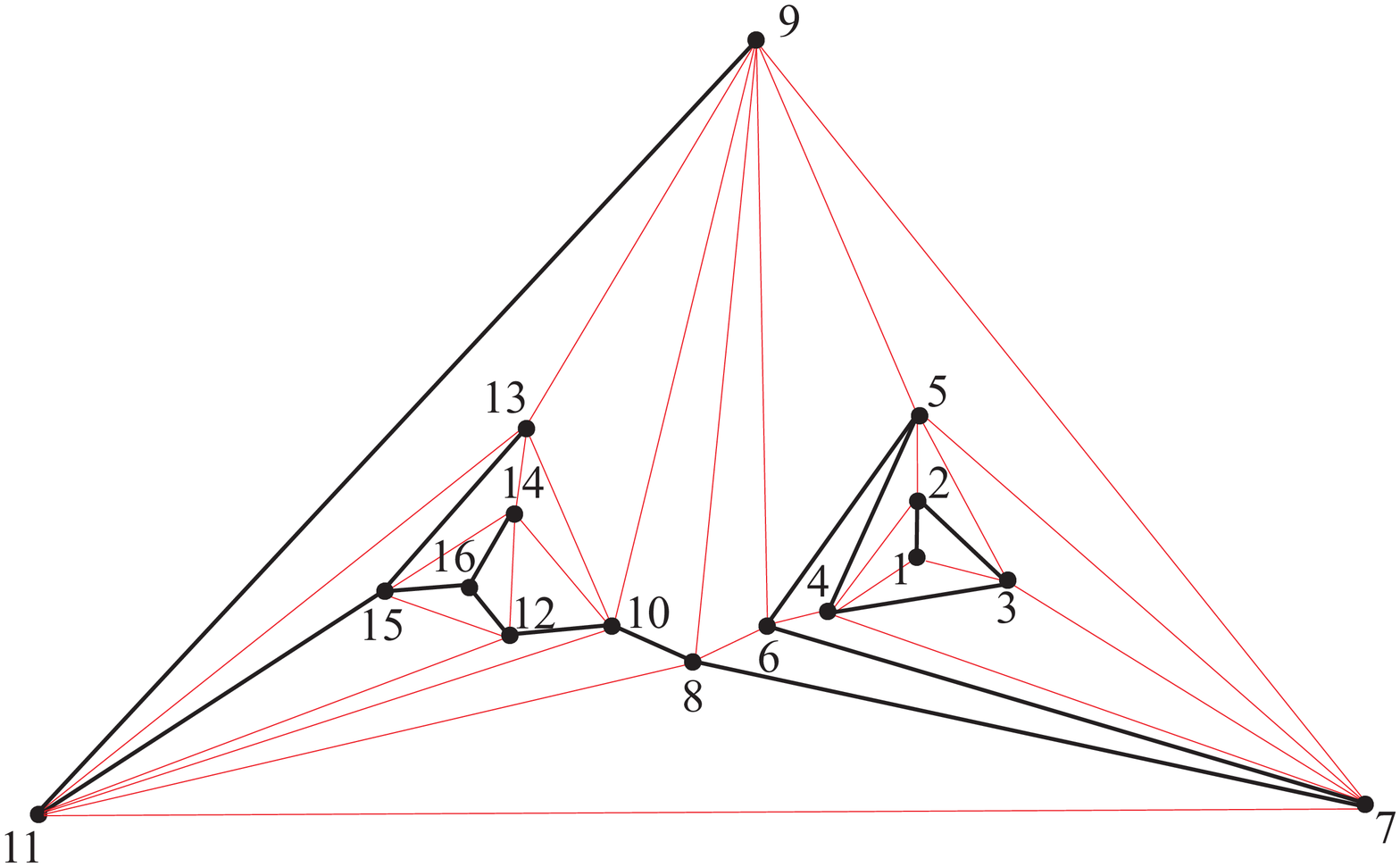}
\end{center}
\caption{Planar representation of the embedding on $S_0$ of the PMFG of the 16 Eurodollar interest rates.
The labels are the same as in Fig.\ref{f.3DEurodollar}.}
\label{f.3DEurodollar1}
\end{figure}

\subsection{PMFG for $16$ Eurodollar interest rates US stocks}

Let us here start with the analysis of the Eurodollar interest rates.
For several economic reasons, interest rates have very similar statistical behaviors and follow similar trends in time. 
This makes the subject very challenging since one is no more dealing with the statistics of single objects but with the notion of a whole complex set of interacting elements which collectively fluctuate~\cite{Pagan,Rebonato,Bouchaud,DiMatteo,DiMatteoMant,DiMatteoScalas,Nuyts}. 
Here we apply our filtering procedure to daily values in the time period $1990-1996$ for $16$ Eurodollars interest rates with maturity date between $3$ to $48$ months \cite{DiMatteo,DiMatteoAless04}. 

These interest rates are very highly correlated with correlation coefficients values between $0.46$ and $0.98$ with $76\%$ of the coefficients larger than $0.8$.
Fig.~\ref{f.MSTEurod} shows the MST associated with this set of data. The labels represent the maturity dates (in units of $3$ months) and the distances between the nodes are set to the corresponding metric distances $d_{i,j}=\sqrt{2(1-c_{i,j})}$. Fig.~\ref{f.3DEurodollar} shows a three dimensional representation of the PMFG network where the distance between nodes is $d_{i,j}$. 
Fig.~\ref{f.3DEurodollar1} reports a drawing of the PMFG network on the Euclidean plane.
In this last figure, the thicker lines indicate links belonging to both the MST and the PMFG. This figure gives an immediate graphical proof that the PMFG network is planar ($g=0$): indeed it can be drawn on the plane without edge-crossings.
In this PMFG we count $38$ cliques of 3 elements (triangles). $28$ of such triangles lie on the planar surface whereas the other $10$ are `collar rings'. 
The total number of possible combinations of $n$ elements in groups of three (i.e. the total number of cliques of $3$ elements in the complete graph) is $\binom{n}{3}=560$. This number is much larger than the number of $3$-cliques in PMFG ($38$). 
Similarly, the number of cliques of $4$ elements (tetrahedra) is $10$, a number which is much smaller than the number of cliques of $4$ elements present in the fully connected graph $\binom{n}{4} = 1820$. 
Indeed, the planar embedding reduces the interconnectivity of the network simplifying the resulting graph. In Table~\ref{t.1} all the cliques of $4$ elements are reported. In this table are also reported the average correlation coefficients $\left<c_{i,j}\right>$ inside each clique, the difference $\Delta$ between the maximum and minimum correlation coefficient and the standard deviation $\sigma$.

\begin{figure}
\begin{center}
\includegraphics[width = 0.5 \textwidth]{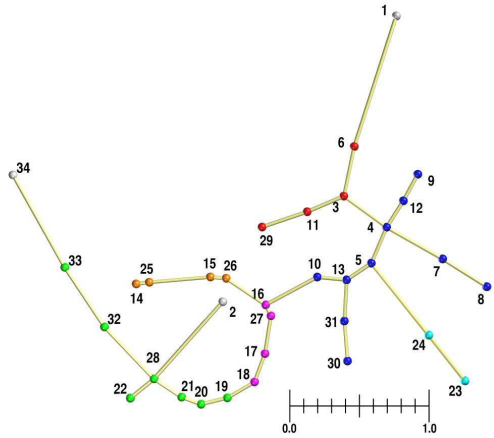}
\end{center}
\caption{Two dimensional representation of the minimum spanning tree for the $34$ US Interest rates. The edge-lengths are equal to the metric distances $d_{i,j}=\sqrt{2(1-c_{i,j})}$. 
The labels represent different interest rates described in the Ref. \cite{DiMatteoMant}.
}
\label{f.MSTutTas}
\end{figure}

\begin{figure}
\begin{center}
\includegraphics[width = 0.5 \textwidth]{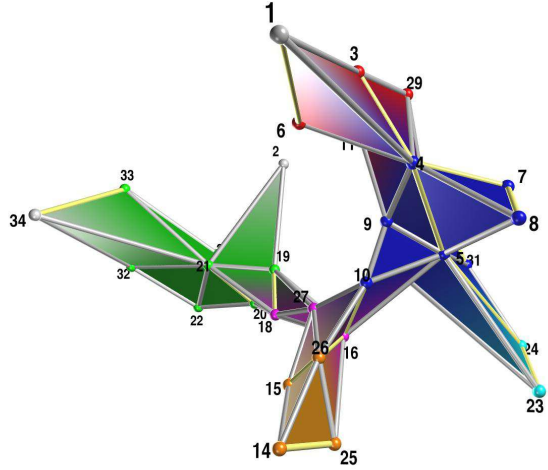}
\end{center}
\caption{Three dimensional representation of the embedding on $S_0$ of the PMFG of the correlation structure for the $34$ US Interest rates. Each edge-length corresponds to the metric distance $d_{i,j}=\sqrt{2(1-c_{i,j})}$.
The labels represent different interest rates described in the Ref. \cite{DiMatteoMant}.
}
\label{f.3DTuttiTassi}
\end{figure}

\begin{figure}
\begin{center}
\includegraphics[width = 0.5 \textwidth]{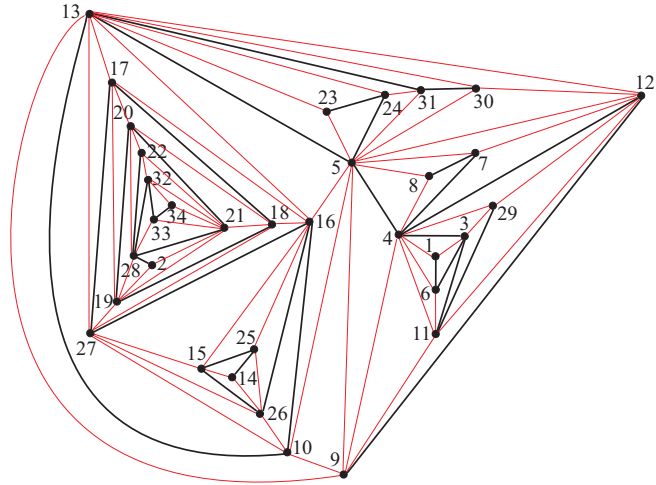}
\end{center}
\caption{Planar representation of the embedding on $S_0$ of the PMFG of the correlation structure for the $34$ US Interest rates.
The labels represent different interest rates described in the Ref. \cite{DiMatteoMant}.
}
\label{f.3DTuttiTassi1}
\end{figure}

\subsection{PMFG for $34$ US interest rates} 

The second example regards the analysis of correlation coefficients among $34$ different weekly interest rates recorded during a time period of $16$ years between $1982$ and $1997$ and stored in the Statistical Release database of the Federal Reserve~\cite{DiMatteoMant,data}. 
For such a data-set we construct both the MST (Fig.~\ref{f.MSTutTas}) and the PMFG (Figs.~\ref{f.3DTuttiTassi} and~\ref{f.3DTuttiTassi1}). 
In this case we observe $94$ cliques of $3$ elements and $31$ cliques of $4$ elements. 
Note that, also in this case, such numbers are much smaller than the number of all possible cliques of $3$- and $4$-elements in $K_{34}$ which are respectively $\binom{n}{3} = 5984$ and $\binom{n}{4} = 46376$.
The complete list of $4$ cliques together with $\left<c_{i,j}\right>$, $\Delta$ and $\sigma$ are listed in Table~\ref{t.2}. 

\begin{table}
\nonumber
\caption{\label{t.1}
The entire set of $4$-cliques in the Eurodollar interest rates. In 
bold are reported the labels corresponding to maturity dates of $1$, 
$2$, $3$ and $4$ years (labels $4$, $8$, $12$, and $16$ 
respectively.)} \begin{center} \begin{tabular}{ccccccc}
\hline
&& Labels & & $\left< c_{i,j} \right>$ & $\Delta$ & $\sigma$ \\
\hline
{\bf 12} & 14 & 15 & {\bf 16} & 0.977 & 0.017 & 0.006 \\
9 & 10 & 11 &13 & 0.972& 0.016 & 0.006\\
{\bf 8} & 9 & 10 &11 & 0.969 & 0.021& 0.008 \\
7 & {\bf 8} & 9 & 11 & 0.965&0.021 & 0.007 \\
6 & 7 & {\bf 8} & 9 & 0.964& 0.026 & 0.009\\
5 & 6 & 7 & 9 & 0.959 & 0.029 & 0.012\\
{\bf 4} & 5 & 6 & 7 & 0.958 &0.044 & 0.016 \\
3 & {\bf 4} & 5 & 7 & 0.94 & 0.06 & 0.02 \\
2 & 3 & {\bf 4} & 5 & 0.91 & 0.13 & 0.05 \\
1 & 2 & 3 & {\bf 4} & 0.83 & 0.30 & 0.11 \\
\hline
\end{tabular}
\end{center}
\end{table}
\begin{table}\nonumber
\caption{\label{t.2}
The entire set of $4$-cliques in the $34$ US Interest rates.} 
\begin{center} 
\begin{tabular}{ccccccc}
\hline
&& Labels & & $\left< c_{i,j} \right>$ & $\Delta$ & $\sigma$ \\
\hline
22 & 28 & 21 & 20 & 0.967 & 0.045 & 0.016\\
21 & 28 & 20 & 19 & 0.97 & 0.06 & 0.02\\
20 & 21 & 19 & 18 & 0.97 & 0.05 & 0.02\\
9 & 12 & 5 & 4 & 0.962 & 0.032 & 0.013\\
18 & 27 & 17 & 16 & 0.96 & 0.07 & 0.03\\
19 & 20 & 18 & 17 & 0.96 & 0.06 & 0.02\\
26 & 27 & 16 & 15 & 0.96 & 0.07 & 0.03\\
10 & 13 & 9 & 5 & 0.957 & 0.040 & 0.018\\
12 & 13 & 9 & 5 & 0.955 & 0.040 & 0.017\\
30 & 31 & 13 & 5 & 0.951 & 0.044 & 0.017\\
19 & 27 & 18 & 17 & 0.95 & 0.09 & 0.03\\
13 & 30 & 12 & 5 & 0.949 & 0.044 & 0.016\\
25 & 26 & 15 & 14 & 0.93 & 0.10 & 0.05\\
13 & 16 & 10 & 5 & 0.93 & 0.11 & 0.05\\
26 & 27 & 16 & 10 & 0.93 & 0.11 & 0.04\\
17 & 27 & 16 & 13 & 0.93 & 0.13 & 0.05\\
16 & 27 & 13 & 10 & 0.93 & 0.11 & 0.05\\
11 & 12 & 9 & 4 & 0.93 & 0.10 & 0.05\\
7 & 12 & 5 & 4 & 0.92 & 0.10 & 0.04\\
11 & 29 & 4 & 3 & 0.92 & 0.07 & 0.03\\
28 & 32 & 22 & 21 & 0.92 & 0.11 & 0.06\\
6 & 11 & 4 & 3 & 0.92 & 0.09 & 0.03\\
25 & 26 & 16 & 15 & 0.92 & 0.17 & 0.06\\
12 & 29 & 11 & 4 & 0.91 & 0.10 & 0.04\\
7 & 8 & 5 & 4 & 0.90 & 0.10 & 0.04\\
24 & 31 & 13 & 5 & 0.87 & 0.21 & 0.10\\
32 & 33 & 28 & 21 & 0.87 & 0.17 & 0.06\\
21 & 28 & 19 & 2 & 0.84 & 0.26 & 0.13\\
23 & 24 & 13 & 5 & 0.80 & 0.30 & 0.12\\
33 & 34 & 32 & 21 & 0.76 & 0.23 & 0.10\\
4 & 6 & 3 & 1 & 0.7 & 0.5 & 0.2\\
\hline
\end{tabular}
\end{center}
\end{table}

\section{Clique structure}
\label{clique} 

The construction algorithm and the topological constraint of the PMFG force each element to participate to at least a clique of $3$ elements. The PMFG is a topological triangulation of the sphere. 
Therefore the {\it triangular rings} ($3$-cliques) are the elementary building blocks of this network and all the possible PMFG networks associated to the correlations between a set of $n$ elements can be generated by exploring the ensemble of planar triangulations with $n$ nodes.  
In a triangulation of the topological sphere the number of triangles on the surface is $2n -4$. However, in the two systems considered (Eurodollars and US interest rates) as well as in the $100$ stock portfolio examined in Ref.~\cite{TumminielloPNAS05}, the number of $3$-cliques is systematically larger than the number of triangles on the surface indicating therefore that there are several triangular rings  which are collar rings and do not lie on the surface.
By observing in details Figs.~\ref{f.3DEurodollar} and~\ref{f.3DTuttiTassi}, we note that these internal rings belong to tetrahedra which pack together some 3-clique by sharing a triangular ring which then becomes a collar ring.
The basic structures in these graphs are the $4$-cliques, which during the formation of the PMFG clusterize together locally at similar correlation values and then connect to each other by following the MST as skeleton structure.
If such $4$ elements cliques are the `emerging building blocks' of the PMFG, then there must be strong relations between their properties and the ones of the system from which they have been generated.
One can verify that indeed the network for the $34$ US interest rates is entirely made by a set of tetrahedra packed together by sharing a triangular face.
Such a packing leads to a structure made of $3 n-8$ cliques of $3$ elements and $n -3$ cliques of $4$ elements. This is exactly what we find in the case of the $34$ US interest rates.
Interestingly, the planar network associated with the $16$ Eurodollar interest rates does not follow the above scheme having $38$ cliques of $3$ elements (instead of $3 n -8 = 40$) and $10$ cliques of $4$ elements (instead of $n-3 = 13$).
By analyzing the network in detail one can see that such a discrepancy comes from the region where interest rates with large maturity dates of $30$-$48$ months gather together.
However, it is clear from Fig.~\ref{f.3DEurodollar} that, apart from this region, the basic structure is also in this case identifiable with a tetrahedral packing.

\begin{center}
\begin{figure}
\begin{center}
\includegraphics[width = 0.5 \textwidth]{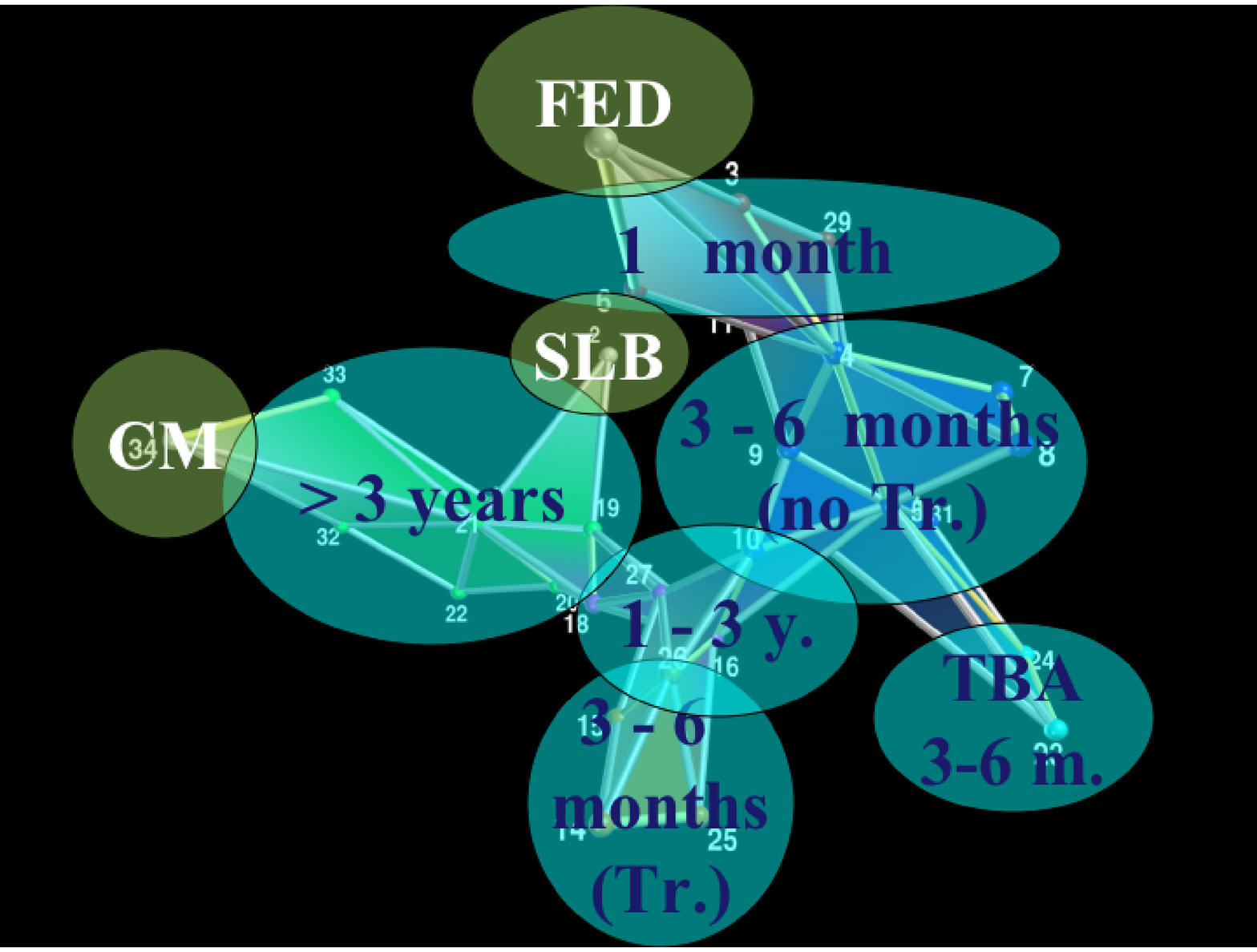}
\end{center}
\caption{\label{f.US_CLU} 
Cluster structure in the PMFG for the $34$ US interest rates.
The structure can be seen as an aggregate of $4$-cliques which reveal the hierarchical organization of the underlying system (see text).
}
\end{figure}
\end{center}

A further analysis of the clique structure for the case of the Eurodollar interest rates shows that these cliques gather together interest rates forming clusters which are characterized by restricted ranges of maturity dates.
From Table~\ref{t.1}, one can note that the most correlated cliques are the ones with largest maturity dates and the spread of the correlation coefficients inside each clique increases as the correlation decreases.

A similar differentiation in term of maturity dates arises also from the analysis of the $34$ US interest rates. In ref.~\cite{DiMatteoMant}, it was shown that these data gather together in $6$ main clusters and three isolated elements. By comparing the structure of $4$-cliques in Table~\ref{t.2} with this cluster gathering we observe that there are $13$ cliques composed by elements all belonging to the same cluster, $11$ cliques with $3$ elements belonging to the same cluster, $5$ cliques which mix two elements for one cluster and two from another and finally $2$ cliques with elements belonging to three different clusters. Fig.\ref{f.US_CLU} shows schematically such a differentiation into different clusters.\\

\section{Conclusions}
\label{conclusion}
The correlation coefficient matrix has been studied in different financial data by analyzing the structure of a network obtained by linking the most correlated elements while constraining the genus of the resulting graph.
We find that in the case $g=0$ corresponding to the PMFG, the basic structure of such network is formed by packing together cliques of $4$ elements which share one or more triangular ring.
The study of such $4$-cliques in the case of $16$ Eurodollar interest rates and $34$ interest rates reveals that the network hierarchy spontaneously generates clusters grouping together interest rates with similar maturity dates. 
This result confirms previous findings~\cite{TumminielloPNAS05} which show such hierarchical gathering and differentiation in the case of $100$ US stocks.

\acknowledgments{
We wish to thank S.T. Hyde and S. Ramsden for fruitful discussions and advices.
We acknowledge partial support from research project MIUR 449/97 ``Dinamica di altissima frequenza nei mercati finanziari''. 
MT and RNM thank partial funding support from research projects MIUR-FIRB RBNE01CW3M, NEST-DYSONET 12911 EU project.
TA and TDM acknowledge partial support from ARC Discovery Project DP0344004 (2003) and DP055813 (2005) and Australian Partnership for Advanced Computing National Facilities (APAC).}
%

\end{document}